# Beyond Keywords and Relevance: A Personalized Ad Retrieval Framework in E-Commerce Sponsored Search

Su Yan
Alibaba Group
yansu.ys@alibaba-inc.com

Wei Lin
Alibaba Group
yangkun.lw@alibaba-inc.com

Tianshu Wu
Alibaba Group
shuke.wts@alibaba-inc.com

Daorui Xiao
Alibaba Group
daorui.xdr@alibaba-inc.com

Xu Zheng
Alibaba Group
zhengxu.zx@alibaba-inc.com

Bo Wu
Alibaba Group
rhope@alibaba-inc.com

Kaipeng Liu
Alibaba Group
zhiping.lkp@taobao.com

## ABSTRACT

On most sponsored search platforms, advertisers bid on some keywords for their advertisements (ads). Given a search request, ad retrieval module rewrites the query into bidding keywords, and uses these keywords as keys to select Top N ads through inverted indexes. In this way, an ad will not be retrieved even if queries are related when the advertiser does not bid on corresponding keywords. Moreover, most ad retrieval approaches regard rewriting and ad-selecting as two separated tasks, and focus on boosting relevance between search queries and ads. Recently, in e-commerce sponsored search more and more personalized information has been introduced, such as user profiles, long-time and real-time clicks. Personalized information makes ad retrieval able to employ more elements (e.g. real-time clicks) as search signals and retrieval keys, however it makes ad retrieval more difficult to measure ads retrieved through different signals. To address these problems, we propose a novel ad retrieval framework beyond keywords and relevance in e-commerce sponsored search. Firstly, we employ historical ad click data to initialize a hierarchical network representing signals, keys and ads, in which personalized information is introduced. Then we train a model on top of the hierarchical network by learning the weights of edges. Finally we select the best edges according to the model, boosting RPM/CTR. Experimental results on our e-commerce platform demonstrate that our ad retrieval framework achieves good performance.

## 1 INTRODUCTION

Sponsored search is a multi-billion dollar industry and is growing dramatically each year. In most sponsored search, paid advertisements (ads) are presented to users along with organic search results. General speaking, sponsored search is composed of three participants. First, advertisers select some keywords for their ads and bid on these selected keywords. Second, users submit search requests, which can be extracted as some intention signals (e.g. queries). Third, sponsored search engines determine which ads to be presented to users, according to the relevance and revenue estimation between search requests and ads. Specifically, in e-commerce sponsored search, the organic search results are trading products named "item", while the ads can be seen as a special kind of promotional products.

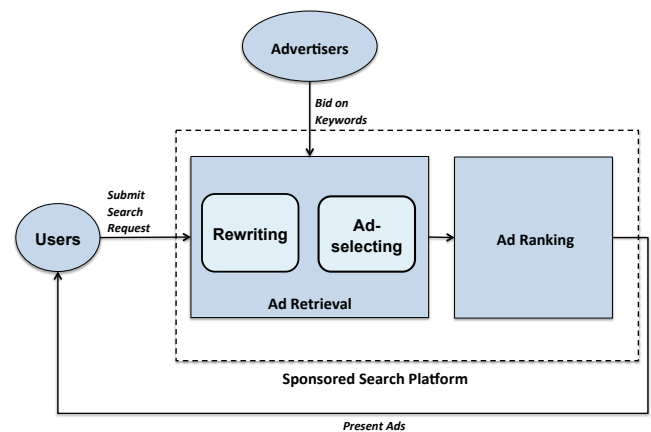

**Figure 1: Sponsored search is composed of three participants: advertisers, users and sponsored search platforms.**

Considering the effectiveness and efficiency, most sponsored search engines consist of two modules: ad retrieval module and ad ranking module. Given a search request, ad retrieval module retrieves a small relevant and high-quality ad collection from a large number of candidates, and then ad ranking module ranks all the retrieved ads by their qualities and bidding prices. Due to the limitation of response time, ad retrieval module only accesses a small amount of information and the computational speed is required to be fast, while ranking module can use more information and employ more complex and time consuming algorithms. This paper mainly focuses on ad retrieval module.

Ad retrieval modules usually contain two traditional tasks: rewriting and ad-selecting. Rewriting is aimed to select a list of keywords, relevant to given search requests/queries. Then all ads bidding on rewritten keywords will be fed into ad-selecting module, which will use these rewritten keywords as keys to select Top N ads through inverted indexes very fast. Considering the efficiency problem, most ad-selecting approaches just employ simple and efficient formulas instead of complex models.

As above, most sponsored search systems require advertisers bid on keywords, and all the retrieved ads are subject to these keywords. An ad will not be retrieved even if search queries are related when the advertiser does not bid on corresponding keywords. However, due to the lack of marketplace information and the cost of management, sometimes advertisers are unable to bid on the best keywords for their ads timely and accurately. In this situation, sponsored search engines cannot achieve the actual global optimal matching, bringing losses to all advertisers, users and sponsored search engines.

Moreover, most existing ad retrieval approaches focus on computing relevance between search queries and ads, and retrieve the most related ads (e.g. [1,2,3]). These approaches are not aimed to boost final Revenue Per Mille (RPM)/Click Through Rate (CTR) performance, and consequently their retrieving results could be sub-optimal.

Recently, on e-commerce sponsored search platforms, more and more personalized information has been introduced, such as user profiles, long-time and real-time clicks. On one hand, by using personalized information, ad retrieval can employ more elements (e.g. real-time clicks) as search intention signals and retrieval keys. These extra personalized signals and keys help ad retrieval better understand users' intentions, and be able to retrieve ads leaving off bidding keywords. On the other hand, personalized information makes ad retrieval a difficult problem to rank ads retrieved through different signals in a uniform way, which is not well solved in most existing ad retrieval approaches.

To address the above problems, we propose a novel ad retrieval framework beyond keywords and relevance in e-commerce sponsored search. Firstly, we employ historical ad click data to initialize a hierarchical network with personalized information, representing multi-relations between user intentions and ads. Specifically, each node in the hierarchical network is tagged as a signal, a key or an ad. The edges between signals and keys stand for rewriting, and the edges between keys and ads stand for ad-selecting. Then we train a model based on the hierarchical network, learning the weights of edges and fine-tuning the network. In this way, the model will combine rewriting with ad-selecting together, and handle all signals from a request at the same time. Finally we select the best edges according to model scores, boosting RPM/CTR. For retrieved ads, there is no bidding information. We use a bid optimizing strategy called Optimized Cost Per Click (OCPC) to determine how much the advertisers will be charged if the ads are clicked.

To summarize, we make the following contributions:

1) Our framework no longer requires advertisers bid on keywords, and uses OCPC strategy to determine ad prices. Hence given a search request, our framework extends the range of candidates, and can achieve better matching.

2) Our framework jointly learns rewriting with ad-selecting together, and is aimed to boost final RPM/CTR, instead of to compute relevance between search requests and ads.

3) Our framework makes use of personalized information to better understand users' intentions, and measures all ad qualities retrieved through different signals in a uniform way.

Experimental results on our e-commerce platform demonstrate that our proposed ad retrieval framework achieves good performance.

## 2 RELATED WORK

Rewriting is a well-studied problem, and a vast amount of methods have been proposed. Most rewriting methods aim to explore the relevance between queries and keywords. Since the text length of queries and keywords is usually very short, *Broder et al.* [1,2] explores the organic search results as additional knowledge features to enrich queries and ads, and built expanded query representations from the preprocessed related queries for real-time rewriting. Similarly, *Choi et al.* [3] uses landing pages to expand the text length. *Jones et al.* [5] employs user query sessions to compute similarities between queries and phrases. By using historical ad click information, *Antonellis et al.* [6] builds a click graph, and then proposes a graph based measure named Simrank++ to identify similar queries. *Hillard et al.* [7] introduces a machine learning approach based on translation models to predict ad relevance, which can help select more relevant ads for the sponsored search system. *Gao et al.* [11, 12] introduces statistical machine translation (SMT) method into query expansion and rewriting problems, and they use sophisticated techniques to avoid sparse data problems. Recently, *Grbovic et al.* [13] applies deep learning techniques and generates distributed language models for queries to improve the relevance in sponsored search, and *Sordoni et al.* [14] proposes a hierarchical recurrent encoder-decoder method for query auto-completion.

Meanwhile, most ad-selecting methods just employ simple and efficient formulas to select Top N ads, due to the limitation of time and space cost. *Broder et al.* [15] proposes the formulation of queries in the format of Weighted AND (WAND) to generate candidates. *McNeill et al.* [17] proposes a dynamic blocking algorithm to choose the blocking keys based on the data characteristics at run time, together with a MapReduce implementation. Some works try to use models to learn candidate selection. *Flake et al.* [16] proposes an approach for constructing query modifications in the web search domain using corpus-based SVM models. *Borisyuk et al.* [18] makes use of WAND query and proposes a machine learned candidate selection framework in LinkedIn's Galene search platform.

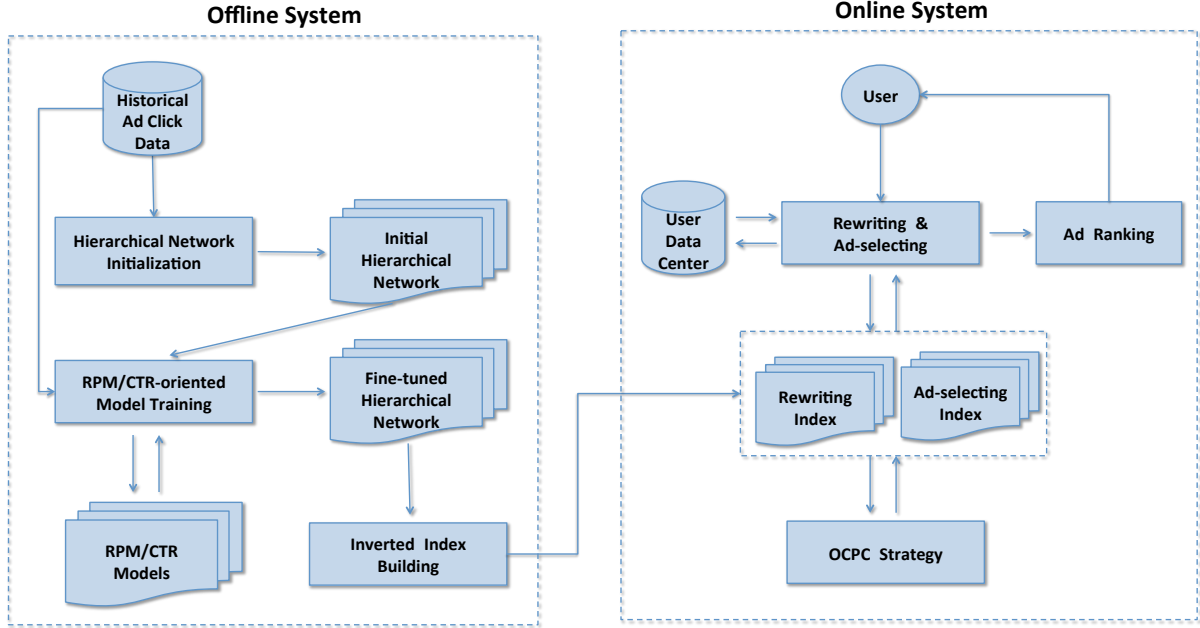

**Figure 2: The architecture of our framework, including offline system and online system. Since this paper focuses on the ad retrieval module, the ad ranking module is simplified in the figure.**

However, all the above approaches regard rewriting and ad-selecting as two separated tasks, and focus on relevance but not final RPM/CTR performance. Only a few of works study this problem. *Malekian et al.* [9] proposes a combinatorial framework for optimizing query rewriting, taking the relevance and budget constraints of ads into account. They regard rewriting as a graph covering problem on a graph of queries, keywords, and ads, and select keywords for each query, so that the benefit of the ads adjacent to the rewrites is maximized subject to ad system and budget constraints. *Cui et al.* [10] treats rewriting as an optimization problem so that rewriting can work together with downstream components. They extract a bunch of features to represent each pair of query and bidding keywords, and train a model to predict whether the word is selected or not, by maximizing the aforementioned marketplace objective. However, the two approaches are quite complex, and difficult to be applied on large-scale online platforms. Moreover, these approaches cannot well solve the personalized multiple signals problem.

Furthermore, all the above approaches require advertisers bid on keywords, and retrieved ads are limited by keywords. Our framework leaves off bidding keywords and use OCPC strategy to determine ad prices. *Zhu et al.* [19] proposes OCPC to optimize advertisers' demands, platform business revenue and user experience and as a whole improves traffic allocation efficiency, and achieves better results than previous fixed bidding manner in Taobao display advertising system in production.

## 3 OUR FRAMEWORK

In this section we present our proposed ad retrieval framework in detail. As mentioned in the introduction, our framework employs personalized information as search intention signals and retrieval keys in e-commerce sponsored search. We no longer require advertisers bid on keywords, and learns rewriting and ad-selecting jointly to boost final RPM/CTR.

Figure 2 shows the architecture of our framework. First we employ historical ad click data to initialize a hierarchical network including signals, keys and ads, in which personalized information is introduced. Then we train a model to mine the most important edges of the hierarchical network, and the model is aimed to boost RPM/CTR. Finally, we build the rewriting index with the edges between signals and keys, and the ad-selecting index with the edges between keys and ads between keys and ad. In this way, we use the two inverted indexes to store the hierarchical network and model scores, retrieving ads within a short time. Since our framework leaves off bidding keywords, we use OCPC strategy to determine how much the advertisers will be charged if the ads are clicked, which can well balance the revenue of advertisers and sponsored search engines.

In Section 3.1 we show how we initialize the hierarchical network, aiming to avoid missing available matches and provide a good starting point for the downstream model training. In Section 3.2 we present our model training method, including studies of features, model types and learning objectives. In Section 3.3 we show how we build indexes and use OCPC strategy to determine ad prices.

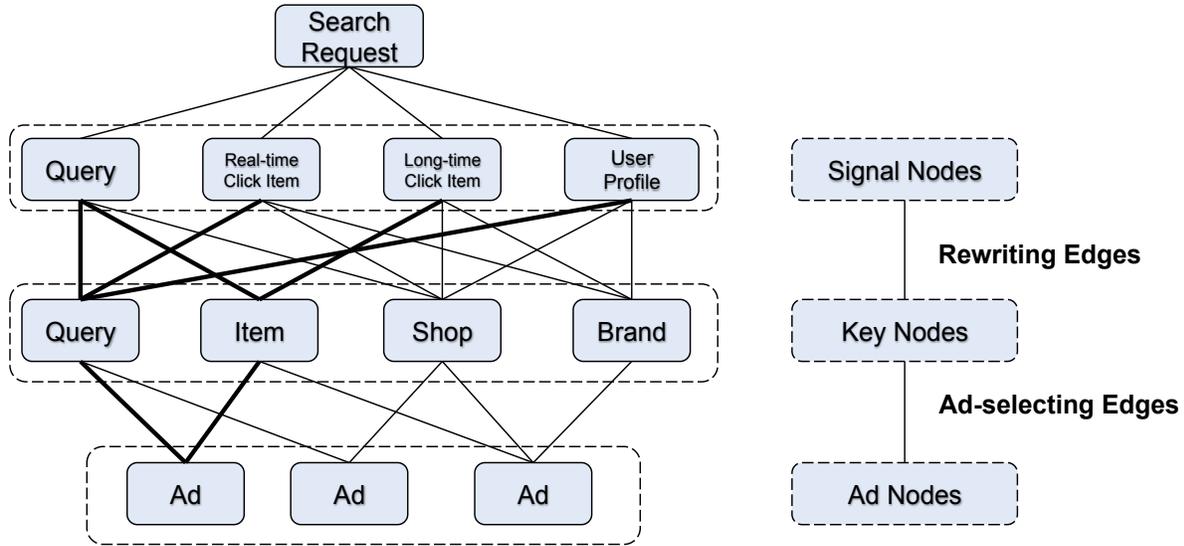

Figure 3: A hierarchical network example with partial nodes and edges. The hierarchical network has three types of nodes: signals, keys and ads. The edges between signals and keys are rewriting edges, and the edges between keys and ads are ad-selecting edges. The bold edges are activated and give an example to show how an ad is reached through the hierarchical network.

## 3.1 Hierarchical Network Initialization

### 3.1.1 Nodes in Hierarchical Network

First we use the historical ad click data to initialize a hierarchical network. As illustrated in Figure 3, the hierarchical network contains three types of nodes: signals, keys and ads.

*Signal Nodes*. The signal nodes in our hierarchical network denote the search intention signals extracted from users' search request. In our framework, we employ personalized information including user profiles, long-time click items, real-time click items and etc. together with search queries as signal nodes. For example, "*Real-time Click Item*" is a kind of personalized signals, denoting the items a user click under the same query. These personalized signals in our hierarchical network help our ad retrieval system better describe users' search requests and better understand users' search intentions. Therefore our ad retrieval framework can mine more information beyond simple and short queries, and achieve better matching performance. In our work, we use "*Query*", "*Real-time Click Item*", "*Long-time Click Item*", "*User Profile*" and etc. as signal nodes..

*Key Nodes*. The key nodes in our hierarchical network denote the ad retrieval keys of ad-selecting inverted index, used to select Top N ads efficiently. In most sponsored search systems, the keys are the keywords that advertisers select and bid on for their ads in advance. However, in our ad retrieval framework, we leave off the bidding keywords and introduce personalized information as key nodes. In some way, these key nodes can be seen as the properties of ads and be used to describe ads. Therefore, by making use of personalized information and historical ad click data, we expect to mine more accurate and suitable "properties" as key nodes to describe ads. In our hierarchical network, when a key node is linked to an ad node, the key can reach the ad and retrieval the ad. Considering the effectiveness and efficiency, the key nodes cannot be too generic to reach too many ad nodes. For example, "*Gender*" and "*Age*" are not good key nodes, because they can reach too many ad nodes, and these ads are unrelated to each other. A good key node should reach a suitable number of ad nodes, and these reached ads should be related and share common properties to well reflect users' search intentions. In our work, we use "*Query*", "*Item*", "*Shop*", "*Brand*" and etc. as key nodes.

*Ad Nodes*. The ad nodes in our hierarchical network denote the ads to be retrieved and presented.

### 3.1.2 Edges in Hierarchical Network

We have two types of edges in the hierarchical network: edges between signals and keys named rewriting edges, and the edges between keys and ads named ad-selecting edges.

*Rewriting Edges*. Rewriting edges are the edges between signal nodes and key nodes, denoting rewriting in ad retrieval. We can rewrite a signal into a key when the signal node is linked to the key node in our hierarchical network. A signal node can be linked to multiple key nodes, while a key node can be linked to multiple signal nodes as well.

*Ad-selecting Edges*. Ad-selecting edges are the edges between key nodes and ad nodes, denoting ad-selecting in ad retrieval. We can select an ad through a key when the ad node is linked to the key node in our hierarchical network. A key node can be linked to multiple ad nodes, while an ad node can be linked to multiple key nodes as well.

### 3.1.3 Hierarchical Network Initialization Methods

The process of hierarchical network initialization is aimed to provide a good starting point for the downstream models to learn and mine important rewriting and ad-selecting edges. So the initial hierarchical network should contain plentiful enough edges to avoid missing available rewriting and ad-selecting edges. Different from the downstream models, our hierarchical network initialization focuses the relevance between nodes based on historical ad click data. In our work, we use three kinds of methods together to initialize the hierarchical network, mining the relevance between nodes from different aspects.

***Click Counts***. The basic idea is to use click counts to reflect the relevance between nodes. First, we use historical ad click data to count the click numbers between key nodes and ad nodes, and remain the edges whose click counts are larger than the threshold. Then we regard the linked key nodes as the properties of ad nodes, and similarly count the click numbers between signal nodes and key nodes. The rewriting edges with large click numbers are as well remained. However, this method is correlated to the presentation times of ads. Ads with more presentation times have a higher probability of large click counts.

***Modified Information Value***. Information Value (IV) [20] is a very useful concept to select important variable. It helps to rank variables on the basis of their importance. For a variable, IV is calculated as the following formula,

$$IV = \sum_i (\frac{Pos_i}{\sum_j Pos_j} - \frac{Neg_i}{\sum_j Neg_j}) In \frac{Pos_i/\sum_j Pos_j}{Neg_i/\sum_j Neg_j} \quad (1)$$

where $i$ denotes the *i-th* sample group that the variable divides, $Pos_i$ denotes the positive sample number of the *i-th* group, and the $Neg_i$ denotes the negative sample number of the *i-th* group. As shown in the formula, IV can balance the coverage and discrimination of a variable.

Because we only care about "positive" edges, we modify the IV formula to calculate the importance of an edge,

$$IV = (\frac{Click}{\sum Click}) In \frac{Click/\sum Click}{Pre/\sum Pre} \quad (2)$$

where $Click$ denotes the click counts of the edge, $\sum Click$ denotes the total click counts, $Pre$ denotes the present counts of the edge, and $\sum Pre$ denotes the total present counts. Then we use Formula (2) to select the edges with high IV, initializing the hierarchical network. In this way, these selected edges can cover a wide variety of ads as well as can reach ads with good qualities.

***Session-based Relevance***. Methods based on sessions are also good at mining the relevance between nodes. On e-commerce platforms, a session is a series of a user's actions taken within a period of time, such as submitting search queries, clicking items and clicking ads. Usually, the actions within a session are relevant to each other. Therefore we can make use of session information to mine the relevance between nodes.

As in illustrated in Figure 4, we reassemble historical ad click data into sessions. In our work, the actions in sessions are "*Submitting Search Query*", "*Clicking Item*", "*Clicking Ad*" and etc. Each action denotes a node in our hierarchical network. Specifically, the category information is used to ensure the actions within a session are real relevant to each other. Then we use the sessions to describe actions, and calculate the relevance between actions with cosine distances. Obtaining the session-based relevance, we select edges and initialize our hierarchical network.

## 3.2 Model Training

In Section 3.1, we initialize a hierarchical network containing three types of nodes and two types of edges. The initial hierarchical network is dense and has plentiful of edges. Moreover, the rewriting and ad-selecting edges in the initial hierarchical network represent the relevance between nodes, but are not RPM/CTR oriented. So we train models based on the initial hierarchical network to learn the weights of the edges, and mine the most important edges to boost RPM/CTR.

The train/test data is also generated from historical ad click data with the form <{*signal*}, *ad*, *label*>, where {*signal*} are the signals extracted from the search request, *ad* is the presented ad, and the *label* is 1 when the ad is clicked else 0. We extend <{*signal*}, *ad*, *label*> into <{*signal->key*}, {*key->ad*}, *label*> by mining *signal->key->ad* paths from our initial hierarchical network, where {*signal->key*} denote the rewriting edges, and {*key->ad*} denote the ad-selecting edges. Thus we obtain the train/test samples and can design features based on the edges.

Compared to ad ranking, ad retrieval faces the crucial challenge of efficiency. The information we can access and use in ad retrieval is quite limited. Specifically, all the features are required to be extracted fast form inverted indexes. Therefore, we design two kinds of features in our work.

***Sparse Features***. One kind of features is the sparse ID features. We assign IDs for each node and edge in our initial hierarchical network, and use these IDs as features directly. This kind of features has a very high dimensionality feature space. They are fine-grained, but weak at generalization.

***Continuous Features***. Another kind of features is the continuous statistic features. We employ various statistic values (e.g. Click Counts, Present Counts, CTR) as the features to describe the edges. These features are low dimensional. Compared to the high-dim sparse ID features, the continuous features help improve the feature coverage and enhance the model stability.

Besides features, we explore some different types of models based on our hierarchical network. Different from ad ranking module, ad retrieval module requires models to be very high-efficiency due to the crucial limitation of time and space cost. The models in ad retrieval module must compute fast, and use little space. In this work, we try three kinds of models: Logistic Regression (LR), Gradient Boosting Decision Tree (GBDT), and Multilayer Perceptron (MLP). LR is a fast linear model, while GBDT and MLP are nonlinear models. Considering the implement of the sponsored search system, we use both sparse ID features and continuous features in LR, and only continuous features in GBDT and MLP. Specifically, in MLP model, we extra perform some nonlinear transformation on the continuous statistic features in advance. This artificially pre-computing helps reduce the layer number of MLP, and balance the effectiveness and efficiency.

Since we generate train/test samples from historical ad click data, the model we train is naturally a CTR-oriented model. Here we present a method to train a RPM-oriented model. The basic idea is to use ad prices to weight the samples. The positive (clicked) samples are weighted by the ad prices, and the negative (unclicked) samples all receive unit weight. In this way, the target of our models is,

$$Score = \frac{CTR \times Price}{(1-CTR) + CTR \times Price} = \frac{RPM}{(1-CTR) + RPM} \quad (3)$$

$$ODDS = \frac{Score}{1-Score} = \frac{RPM}{1-CTR} \approx RPM \quad (4)$$

Specifically, when two ads have the same RPM, our RPM-oriented models prefer the one with a higher CTR.

With the RPM/CTR-oriented models, we can select the most important edges in initial hierarchical network to boost RPM/CTR. The hierarchical network is fine-tuned by the models. Then given a search request, signals are extracted from the request and sent into the hierarchical network. All the signals will pass through the hierarchical network via the rewriting and ad-selecting edges that our RPM/CTR-oriented models mine, and reach a collection of ad nodes simultaneously. Specifically, an ad can be reached through different paths at the same time. For each reached ad, our model will predict the RPM/CTR based on the activated edges, and determine where the ad should be retrieved or not.

## 3.3 Indexes Building and OCPC Strategy

### 3.3.1 Two Inverted Indexes

With the hierarchical network and RPM/CTR-oriented models, we can select the most important edges and build two inverted indexes to store them. The two inverted indexes are the rewriting index and the ad-selecting index.

***Rewriting Index***. We build the rewriting index to store the high scored rewriting edges that the RPM/CTR-oriented models mines. The triggers of the rewriting index are signal nodes in the hierarchical network, and the terms are key nodes. If a rewriting edge *signal->key* is selected by our models, the corresponding key will be mounted under the corresponding signal. The edge weights and features are stored in the inverted index, in order to rank ads fast.

***Ad-selecting Index***. Similarly, we build the ad-selecting index to store the high scored ad-selecting edges. The triggers of the ad-selecting index are key nodes in the hierarchical network, and the terms are ad nodes. The edge weights and features are also stored in the index.

Through the rewriting index and ad-selecting index, we can retrieve ads fast for online search requests. With the edge weights and features stored in the indexes, we rank all retrieved ads orienting RPM/CTR. For LR model, we simply sum the weights of activated edges to obtain the ranking score. For GBDT model, we encode the model into conditional statements such as "*if*" and "*else*" with edge weights and features for fast computing. For MLP model, we store weight matrixes in memory system.

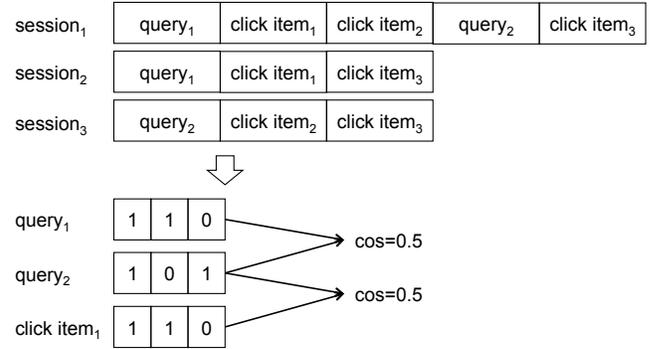

**Figure 4: A simple example for session-based relevance**.

### 3.3.2 OCPC strategy

Since there is no bidding information for retrieved ads in our framework, we use the OCPC strategy to determine how much advertisers will be charged if their ads are clicked.

On most e-commerce sponsored search platforms, the advertisers are precisely the sellers on the same platforms. These sellers advertise with a percentage (taking rate) of their revenue for marketing purposes, and use indices such as Conversion Rate (CVR) to measure traffic values. More valuable traffic brings more revenue, and advertisers are willing to pay higher ad prices. Based on the unique ecosystem characteristics of e-commerce sponsored search platforms, the OCPC is calculated to measure traffic values,

$$OCPC = CVR \times Item\ Price \times Taking\ Rate \quad (5)$$

where CVR is $Trade\ Counts/Click\ Counts$, *Item Price* denotes the selling price of items, and *Taking Rate* denotes the percentage that advertisers are willing to pay for purposes (e.g. 0.05). Our framework estimates CVR and *Taking Rate* by using historical data.

As shown in Formula (5), the OCPC strategy aims to adjust ad prices toward the true value of traffic. The strategy well balances the gain and risk between advertisers and e-commerce sponsored search platforms. On one hand, the Return On Investment (ROI) of advertisers is reasonable and acceptable, based on the product qualities and attractions to users. One the other hand, the sponsored search platforms still take the responsibility to bring good matching between search requests and ads. Previous experimental results on our e-commerce platform have demonstrated that the OCPC strategy can achieve better results than fixed bidding manner.

## 4 EXPERIMENTAL EVALUATIONS

In this section we evaluate the performance of our ad retrieval framework through offline and online experiments. First, in Section 4.1 and 4.2, we conduct offline experiments to evaluate and compare the explorations of model training. Then in Section 4.3 and 4.4, we conduct online experiments on our e-commerce platform to see the actual results of our new ad retrieval framework.

## 4.1 Offline Datasets

We use the historical ad click data on our e-commerce platform to generate the datasets. Table 1 and Table 2 give a brief summary of our datasets.

Table 1: Summary of Hierarchical Network

|  | Counts (order of magnitude) |
|---|---|
| Sessions | $10^{10}$ |
| Signal Nodes | $10^{10}$ |
| Key Nodes | $10^{10}$ |
| Ad Nodes | $10^{7}$ |
| Rewriting Edges | $10^{12}$ |
| Ad-selecting Edges | $10^{11}$ |

As shown in Table 1, we sample approximately tens of billions sessions to initialize our hierarchical network. The initial hierarchical network contains about tens of billions signal nodes, tens of billions key nodes and tens of millions ad nodes. The number of edges between signals and keys is hundreds of billions, and the number of edges between keys and ads is also hundreds of billions. General speaking, the initial hierarchical network is huge and contains plentiful enough edges.

Table 2: Summary of Datasets

|  | Data Days | Sample Number |
|---|---|---|
| Train Data for ID Features | 7 Days | $10^{10}$ |
|  | 28 Days | $10^{10}$ |
|  | 63 Days | $10^{11}$ |
| Test Data for ID Features | 7 Days | $10^{8}$ |
|  | 28 Days | $10^{9}$ |
|  | 63 Days | $10^{9}$ |
| Train Data for Continuous Features | 1 Days | $10^{9}$ |
| Test Data for Continuous Feature | 1 Days | $10^{7}$ |
| Next Data for Both | 1 Days | $10^{9}$ |

As shown in Table 2, we generate three kinds of datasets for model training and testing. The train datasets and test datasets are sampled together from the same period of historical ad click data, and the size of test datasets is about five percent of the size of train datasets. The next dataset is sampled from the next period of historical ad click data, in order to evaluate our models with Out-of-Time (OOT) results. For models with the high-dim sparse ID features, the size of train dataset should be large enough to avoid the overfitting problem, while for models with the low-dim continuous statistic features, the size of train dataset could be much smaller.

## 4.2 Offline Metrics and Experimental Results

First we conduct offline experiments to evaluate the rankings provided by each model. We use the popular metric the area under the ROC curve (AUC) for evaluation,

$$AUC = \frac{\sum_{i \in Pos} Rank_i - \frac{M \times (M+1)}{2}}{M \times N} \quad (5)$$

where $Rank_i$ denotes the ranking of $i$-th positive sample in the model, $M$ denotes the number of positive samples, and $N$ denotes the number of negative samples. The AUC metric is a value between zero and one, and is the larger the better.

A series of experiments are conducted. We use these experiments to explore the effects of different types of features and different types of models mentioned in Section 3.2. Table 3 shows the comparison results of each model.

Table 3: Offline Results for Models

|  | Train AUC | Test AUC | Next AUC |
|---|---|---|---|
| LR_ID_7 | 0.656 | 0.636 | 0.630 |
| LR_ID_28 | 0.660 | 0.652 | 0.649 |
| LR_ID_63 | 0.690 | 0.665 | **0.658** |
| LR_CON | 0.647 | 0.647 | 0.647 |
| GBDT_CON_30 | 0.662 | 0.662 | 0.661 |
| GBDT_CON_100 | 0.664 | 0.664 | **0.664** |
| MLP_CON_LIN | 0.657 | 0.657 | 0.657 |
| MLP_CON_NONLIN | 0.661 | 0.661 | **0.660** |

.

*LR_ID_7*. A LR model with sparse ID features. The train data is the 7-days train dataset.

*LR_ID_28*. A LR model with sparse ID features. The train data is the 28-days train dataset.

*LR_ID_63*. A LR model with sparse ID features. The train data is the 63-days train dataset.

*LR_CON*. A LR model with continuous features. The train data is the 1-day train dataset. The continuous features are counted from a 63-days dataset.

*GBDT_CON_30*. A GBDT model with continuous features, containing 30 tress. The train data is the 1-day train dataset. The continuous features are counted from a 63-days dataset.

*GBDT_CON_100*. A GBDT model with continuous features similar to GBDT_CON_30. GBDT_CON_100 contains 100 tress.

*MLP_CON_LIN*. A MLP model with continuous features. The model is a 3-layer full connected network. The train data is the 1-day train dataset. The continuous features are counted from a 63-days dataset.

*MLP_CON_NONLIN*. A MLP model similar to MLP_CON_LIN. We perform some nonlinear transformation to extend these continuous features in advance.

As seen from Table 3, the results of models with sparse ID features (LR_ID_7, LR_ID_28 and LR_ID_63) demonstrate that for the high-dim sparse ID features, the size of training dataset is very important. With more samples being fed into, the models can achieve better performance, and the result gap among train data, test data and next data will be smaller. This is easy to understand because the high-dim ID features are quite fine-grained features, and are weak at generalization. Therefore the models with ID features are just sample-memory models. The more samples these models see, the better performance they can achieve.

Meanwhile, the results of models with continuous statistic features demonstrate that for the low-dim continuous statistic features, the training dataset can be much smaller. Their results on train data, test data and next data are quite stable, which is consistent with our expectation.

The results in Table 3 also show that for continuous statistic features, GBDT and MLP perform better than LR. This is because some of the statistic features are not linear correlated to RPM/CTR. GBDT and MLP can mine nonlinear patterns from train data, which help improve the results.

### 4.3 Online Indexes

In online system, the sizes of indexes are limited considering time and space cost. For the rewriting index, the number of triggers related to one signal is strictly limited to small size (e.g. 100), as search in retrieval index for too many triggers is time consuming. For the retrieval index, the number of ads related to one trigger is limited (e.g. no more than 300) with primary consideration of the machine memory.

Table 4: Online Rewriting Index

|  | Number (order of magnitude) |
|---|---|
| Total Size | $10^{10}$ |
| Trigger Size | $10^{10}$ |
| Average Length | $10^{2}$ |
| Median Length | $10^{2}$ |

Table 5: Online Ad-selecting Index

|  | Number (order of magnitude) |
|---|---|
| Total Size | $10^{10}$ |
| Trigger Size | $10^{10}$ |
| Average Length | $10^{2}$ |
| Median Length | $10^{2}$ |

Table 4 and Table 5 give a summary of the indexes we use in this paper for experiments. Through the rewriting index and ad-selecting index, we can retrieve the top RPM /CTR ads fast for online search requests.

### 4.4 Online Metrics and Experimental Results

Then we conduct online experiments on our platform with real traffic. We use three online metrics to evaluate the performance of our ad retrieval framework.

*Click Through Rate (CTR)*:

$$CTR = Click\ Counts/Present\ Counts$$

*Revenue Per Mille (RPM)*:

$$RPM = CTR \times Ad\ Price$$

*Present Rate (PR)*:

$$PR = \mathrm{I}(Present\ Counts > 0)/Request\ Counts$$

where $\mathrm{I}(Present\ Counts > 0)$ is 1 when *Present Counts* > 0 else 0.

Considering the effectiveness and efficiency, we use the model of LR-ID-63 for online experiments. We evaluate LR-ID-63 on 1% traffic of our e-commerce platform. The baseline is a graph covering method similar to [9], which has been shown very effective on our platform, outperforming lots of conventionally incorporated methods such as Simrank++ in [6]. Table 6 summarizes the average metric lift rates.

Table 6: Average Metric Lift Rates on Online Traffic

|  | Lift Rate |
|---|---|
| CTR | 2.0% |
| RPM | 8.0% |
| PR | 1.2% |

As is shown in Table 6, the CTR metric increases 2.0%, and the PR metric increases 1.2%. These two metrics demonstrate that our proposed ad retrieval framework has a better understanding of the users' search intentions, and achieves better matches between search requests and ads. The lifts mean advertisers can obtain better traffic for their ads. The lifts also bring values to search users, who can spend less time to find the items they are really interested in.

Meanwhile, we can see the RPM metric increases 8.0%, showing that our new ad retrieval framework brings values to the sponsored search platform, which can gain revenue significantly.

## 4 CONCLUSIONS

In this paper, we propose a novel ad retrieval framework beyond keywords and relevance in e-commerce sponsored search. Our proposed framework introduces personalized information into ad retrieval to better understand users' search requests and achieve better matches of search requests and ads. We employ historical ad click data to initialize a hierarchical network and train models based on the hierarchical network to boost RPM/CTR. We use two inverted indexes to store the hierarchical network and the RPM/CTR-oriented models, in order to retrieve ads efficiently for online system. OCPC strategy is used to determine how much the advertisers will be charged if the ads are clicked.

Our proposed framework no longer requires advertisers bid on keywords, and focus on final RPM/CTR instead of relevance. We jointly learn rewriting with ad-selecting together. Moreover, our framework well solve the multiple personalized signal problem and measures all ad qualities retrieved through different signals in a uniform way.

We conduct offline and online experiments to evaluate our proposed framework. Actual results on our e-commerce platform demonstrate that our ad retrieval framework achieves good performance, bringing values to all advertisers, users and sponsored search engines.